\documentclass[12pt]{article}
\begin{document}
~
~
~
~
\begin{center} {\large \bf  Gravitational Field of Radiating Cosmic String}

\vspace{1cm}

                      Wung-Hong Huang\\
                       Department of Physics\\
                       National Cheng Kung University\\
                       Tainan,\, Taiwan\\
\end{center}
\vspace{1cm}
We examine the Einstein equation with the energy-momentum tensors which 
correspond to an infinite line string of finite radius plus an outgoing radiation field. It is used to see the effect of the radiation field on the spacetime of a cosmic string. We make some assumptions about the metric coefficients and find a
class of exact solution. The result can be applied to study the back reaction on a radiating cosmic string.
\vspace{3cm}
\begin{flushleft}

Physical Review D47(1993)711

E-mail:  whhwung@mail.ncku.edu.tw\\
\end{flushleft}
\newpage
Cosmic strings may have been created during the phase transition in the early Universe [1]. They could be seeds for galaxy formation [2] and could act as a gravitational lens [3]. Many discussions about cosmic-string physics [4] have been based on the model of a string as a vacuum exterior and nonvacuum interior solution of Einstein's equation with cylindric symmetry [3,5,6]. Both the static
[3,5] and nonstatic [6] solutions have been investigated.

   It is well known that particles can be created from a nonstatic background spacetime [7]. The problem may be treated by calculating the Bogoliubov transformation associated with the changing gravitation field which may be isotropic [8], anisotropic [9], or inhomogeneous [10].  Parker [11] has recently considered the creation of massless scalar field particles minimally coupled to the scalar curvature in a spacetime in which the initial geometry was described by the Minkowski metric, and the final geometry was described by the conical cosmic-string metric [3]. The problem has also been investigated by other authors in a more general situation [12]. However, it is fair to say that, as the particles are created from the string the spacetime outside the string thus was not a vacuum, the geometry they used, which is described by the Vilenkin form [3] with a time-dependent conical metric, will not be realized.

   In this paper we will attempt to find the geometry which described an infinite line string with a finite radius and an outgoing radiation field. We hope that such a metric will be more suitable to study the gravitational particle production from a nonstationary cosmic string, especially to studying the effects of back reaction on a radiating cosmic string. Note that this approach is like those using the Vaidya [13], Vaidya-Bonnor [14], or Vaidya-de Sitter [15] metric to investigate Hawking radiation in a quantum black hole [16-18].

  We start with a general cylindrically symmetric metric [19]
     $$ds^2= - e^{2( K - U)}\left( dt^2 - dr^2 \right) +  e^{-2U}\,W^2\, d\varphi^2+ e^{2U} dz^2, \eqno{(1)}$$ 
where U, K, and W are functions of t and r, $\varphi$ is the azimuthal coordinate, and z is the axial coordinate. The energy-momentum tensor of a string is chosen to be [3,6] 
$$T^\nu_{\mu (string)} = - \sigma (t, r)\,diag(1,0,0,1), \eqno{(2)}$$
while that of a pure radiation field is
$$T^\nu_{\mu (radiation)} = \phi (t,r) \,K_\mu K^\nu ,~~~~ K_\mu K^\nu =0 , \eqno{(3)}$$
where $K_\mu$ is a radial null vector field and $\phi$ is related to the energy density of an outgoing radiation. Define the total energy momentum
$$T^\nu_{\mu} = T^\nu_{\mu (string)} + T^\nu_{\mu (radiation)} , \eqno{(4)}$$
then in the coordinate basis (1) we have $K_\mu =  (1,1,0,0)$ and the nonvanishing components of the total energy momentum become
$$T^t_t = -\sigma - \phi ,$$
$$T^r_r = T^r_t= -T^t_r = \phi  , \eqno{(5)}$$
$$T^z_z = -\sigma .$$
The Einstein field equations ($G_\mu^\nu = 8\pi \, T_\mu^\nu$) to be solved are
$$e^{2(U-K)}\left[-{W^{''} \over  W} + \dot K {\dot W \over  W} +  K^{'} {W^{'} \over  W} -\dot U ^2 - U^{'2} \right]  = 8\pi (\sigma + \phi), \eqno{(6)}$$
$$e^{2(U-K)}\left[{\ddot W \over  W} - \dot K {\dot W \over  W} -  K^{'} {W^{'} \over  W} +\dot U ^2 + U^{'2} \right]  =  - 8\pi  \phi, \eqno{(7)}$$
$$\ddot K - K^{''} + \dot U^2 - U^{'2} = 0 , \eqno{(8)}$$
$$e^{2(U-K)}\left[\ddot K - K^{''} - 2\ddot U +2  U^{''}+  {\ddot W \over  W} - { W^{''} \over W} - 2 \dot U {\dot W \over  W} + 2  U^{'} {W^{'} \over  W} +\dot U ^2 - U^{'2} \right]  = 8\pi  \sigma, \eqno{(9)}$$
$$e^{2(U-K)}\left[\dot K {W^{'} \over  W} +  K^{'} {\dot W \over  W} -{\dot W^{'} \over  W} - 2  \dot U  U^{'} \right]  =  8 \pi \phi, \eqno{(10)}$$
equations (6) and (7), (8)-(11), and (7) and (10) will imply the following equations, respectively:                            
$$e^{2(U-K)}\left[{\ddot W  \over  W} - {W^{''} \over  W} \right]  =  8 \pi \sigma, \eqno{(11)}$$
$$\ddot U - U^{''} +  \dot U {\dot W \over  W} -  U^{'} {W^{'} \over  W}  = 0, \eqno{(12)}$$
$${\ddot W \over  W} - { \dot W^{'} \over W} + (\dot U- U^{'})^2 - (\dot K - K^{'}) \left[ {\dot W \over  W} - { W^{'} \over W}\right]  = 0 . \eqno{(13)}$$
The conservation equations ($T^\nu_{\mu ;\nu} = 0$)  give the relations 
$$\dot \sigma  + \sigma \left[\dot K - 2 \dot U + {\dot W \over  W}\right] + \dot \phi - \phi ^{'} - \phi \left[ 2(\dot U - \dot K) - 2(U^{'}- K^{'})- \left[ {\dot W \over W} - {W^{'} \over W}\right] \right] = 0 . \eqno{(14)}$$
$$\dot \phi - \phi^{'} - \phi \left[ 2(\dot U - \dot K) - 2(U^{'}- K^{'})- \left[ {\dot W \over W} - {W^{'} \over W}\right] \right] - K^{'}\sigma = 0 . \eqno{(15)}$$
Equations (14) and (15) imply
$$\dot \sigma  + \sigma\left[\dot K - 2 \dot U + {\dot W \over  W}+  K^{'}\right] = 0 . \eqno{(16)}$$
Substituting  Eqs. (10) and (11) into Eq. (15), and substituting Eq. (11) into Eq. (16) give the following
$$2 \left[{\dot W \over W} - {W^{'} \over W} \right]\dot U U^{'} + 2 \left[\ddot U U^{'} + \dot U \dot U^{'} - \dot U U^{''}-\dot U^{'} U^{'}\right] + {\ddot W^{'} \over W} - {\dot W^{''}\over W}+ (K^{''}-\dot K^{'}){\dot W\over W}$$
$$~~~~~~~~~~~~~~~ - (\dot K - K^{'}) \left[{\dot W^{'}\over W} - {W^{''} \over W} \right] - (\ddot K - \dot K^{'}) \,  {W^{'}\over W} = 0 ,  \eqno{(17)}$$
$$ {\ddot W^{'} \over W} - {\dot W^{''}\over W}- (\dot K - K^{'}) \left[{\ddot W\over W} - {W^{''} \over W} \right] = 0 ,  \eqno{(18)}$$
Henceforth we shall solve Eqs. (8), (12), (13), (17), and (18), and then use Eq.(10) to evaluate $\phi$ and Eq. (11) to evaluate $\sigma$.

  It seems that without further assumption about the metric forms one is unable to find any analytic solution. In [6], Shaver and Lake found all possible nonstationary solutions of the above equations in the case of  $\phi = 0$ under
the simplifying assumption that the metric coefficients are  separable  functions  of  their  arguments,  i.e., $ U= u(t) + \mu (r), ~~W= w(t) \, \Omega(r)$. [Note that if $\phi =0$ then Eq.(15) implies $K' =0$.] We have made efforts to find the analytic solutions with the physical properties of $\sigma > 0 ( =0)$ within (outside) the string and $\phi > 0$ outside the string, which presents a positive outgoing flux, under the assumption of separability of the metric coefficients.   However, the results are just those obtained from the following analysis.

  One can easily see that in the case of
$$U = U_1(t +r) + U_0 ,  \eqno{(19)}$$
$$K = K_1(t +r) + K_0 ,  \eqno{(20)}$$
where $U_i$ and $K_i$ are constants, the equations to be solved become the very simple partial differential equations of  $W$. The equations depend on whether or not $U_1 =0$.

  Case 1. $U_1 \not=0$.  In this case, the equations to be solved are
$$\dot W = W^{'}, ~~~  \ddot W = W^{''} , ~~~~~ outside~ string ,            \eqno{(21)}$$
$$\dot W = W^{'}, ~~~  \ddot W^{'} = \dot W^{''} , ~~~  \ddot W \not= W^{''},~~~~~ inside~ string ,         \eqno{(22)}$$
The general solutions of Eq. (21) are any function form of  $F(t + r)$; however, no consistent solution could be found to satisfy Eq. (22).

   Case 2. $U_1 =0$. In this case, the equations to be solved are
$$\ddot W = \dot W^{'} = W^{''} , ~~~~~ outside~ string ,            \eqno{(23)}$$
$$\ddot W = W^{''}, ~~~  \ddot W^{'} = \dot W^{''} , ~~~  \ddot W \not= W^{''},~~~~~ inside~ string ,         \eqno{(24)}$$
The general solutions of the above equations are
$$ W = a_2 \,(t +r)^2 + a_1\,t +b_1\,r + W_0 ,  ~~~~~ outside~ string ,            \eqno{(25)}$$
$$ W = a_2\,t^2+2a_2\, r\,t  + a_1\,t + f (r)+ W_0, ~~f(r)\not= a_2\,r^2 ,~~~~~ inside~ string ,         \eqno{(26)}$$
where $a_1 ,b_1$ are constants and $f(r)$ is any function of $r$ (but not the form of $a_2\,r^2$), which shall satisfy the match conditions at string radius $r_0$:
$$f (r_0) = b_1\,r_0, ~~~ f ^{'}(r_0) = b_1\,r_0 .     \eqno{(27)}$$
Because of the freedom of  $f(r)$ the match condition can easily be satisfied.  We present the following example which may be the most simple solution of all:
$$ U =U_0,~~ K=K_1(t +r) + K_0 ,$$
$$ W_{outside}= a_l\,t +b_l\,t + W_0 ,        \eqno{(28)}$$
$$ W_{inside}= a_1\,t + sin(r) + W_0 ,      $$
$$K_1 \,{a_1 + b_1 \over a_1 t+ b_1 r + W_0} \, e^{2[U_0 - K_1(t+r) -K_0]} = 8 \pi \phi_{outside}, \eqno{(29)}$$
$$K_1 \,{a_1 - cos(r) \over a_1 t + sin (r) + W_0} \, e^{2[U_0 - K_1(t+r) -K_0]} = 8 \pi \phi_{inside}, \eqno{(30)}$$
$${sin(r) \over a_1 t + sin (r) + W_0} \, e^{2[U_0 - K_1(t+r) -K_0]} = 8 \pi \sigma_{inside}, \eqno{(31)}$$
The match conditions Eq. (27), and the requirements of  $\phi_{outside} > 0$  (which represent a positive outgoing particle field flux) and $\sigma_{inside}$ give the constrains
$$ r_0 = tan(r_0),~~ b_1=cos (r_0),~~  K_1 >0, ~~a_1,\,W_0 \geq 0 .   \eqno{(32)}$$
If  $K_1 =a_1 =0$ then the above solution becomes a static one found previously [3,5].  If $K_1= 0$ then $\phi = 0$ (but $\sigma \not=0$), and Eqs. (28)-(31) represent a new nonradiating solution which was not found in Ref. [6], as our $W$ function is a separable function form of $w(t)+ f(t)$ instead of
the form of  $w(t)\,f(r)$ used in Ref. [6]. Note that many other analytic solutions could be found in the case of the other function form of $f(r)$ and/or $a_2 \not=0$.

   In conclusion, we have examined the Einstein equation with the energy-momentum tensors which correspond to an infinite line string of finite radius plus an outgoing radiation field.  We make some assumptions about the metric coefficients and have found a class of exact solutions. We hope that such a metric will be more suitable to studying the gravitational particle production from a nonstationary cosmic string, especially to studying the back reaction on the radiating cosmic string.
\newpage
\begin{center} {\large \bf  References} \end{center}
\begin{enumerate}
\item T. W. B. Kibble, J. Phys. A 9, 1387 (1976).
\item Ya. B. Zeldovich, Mon. Not. R. Astron. Soc. 192, 663 (1980).
\item  A. Vilenkin, Phys. Rev. D 23, 852 (1981).
\item  A. Vilenkin, Phys. Rep. 121, 263 (1985).
\item  W. A. Hiscock, Phys. Rev. D 31, 3288 (1985); J. Gott, Astrophys. J. 299, 422 (1985).
\item J. A. Stein-Schabes, Phys. Rev. D 33, 3545 (1986); E. Shaver and K. Lake, ibid. 40, 3287 (1989).
\item N. D. Birrell and P. C. W. Davies, "Quantum Fields in Curved Space" (Cambridge University Press, Cambridge, England, 1982).
\item  L. Parker, Phys. Rev. Lett. 21, 562 (1968).
\item Wung-Hong Huang, "Particle Creation in Kaluza-Klein Cosmology", Phys. Lett. A 140 (1989) 280 [gr-qc/0308086].
\item Wung-Hong Huang, "Spectrum of  Particles Created in Inhomogeneous Spacetimes" Phys. Lett. B 244 (1990) 378 [hep-th/0408034] .
\item  L. Parker, Phys. Rev. Lett. 59, 1369 (1987).
\item G. Mendell and W. A. Hiscock, Phys. Rev. D 40, 282 (1989).
\item P. C. Vaidya, Proc. Ind. Acad. Sci. A 33, 264 (1951).
\item W. B. Bonner and P. C. Vaidya, Gen. Relativ. Gravit. 1, 127 (1970).
\item R. L. Mallett, Phys. Rev. D 31,416 (1985); 33, 2201 (1986).
\item W. A. Hiscock, Phys. Rev. D 23, 2813 ( 1981).
\item Y. Kaminaga, Class. Quantum Grav. 7, 1135 (1990).
\item Wung-Hong Huang, "Hawking Radiation of a Quantum Black Hole in an Inflationary Universe", Class. Quantum Grav. 9 (1992) 1199.
\item D. Kramer, H. Stephani, M. MacCallum, and E. Herlt, "Exact Solution of Einstein's Field Equations" (Cambridge University Press, Cambridge, England, 1980).
\end{enumerate}
\end{document}